%% file: acl_format_used.tex
\pdfoutput=1

\documentclass[11pt]{article}

\usepackage[]{ACL2023}

\usepackage{times}
\usepackage{latexsym}

\usepackage[T1]{fontenc}

\usepackage[utf8]{inputenc}

\usepackage{microtype}

\usepackage{inconsolata}

\title{Evaluating Saliency Explanations in NLP by Crowdsourcing}

\author{Xiaotian Lu\thanks{~~Corresponding author. }$~\hspace{0.5mm}^1$ Jiyi Li $^2$ Zhen Wan $^1$ Xiaofeng Lin $^1$ Koh Takeuchi $^1$ Hisashi Kashima$^1$ \vspace{1mm}\\
$^1$Kyoto University \hspace{5mm}$^2$University of Yamanashi\\
{\small $^1$\texttt{{\{lu@ml.ist.,zhenwan@nlp.ist.,lxf@ml.ist.,takeuchi@,kashima@\}}i.kyoto-u.ac.jp}} \\
{\small $^2$\texttt{jyli@yamanashi.ac.jp}}
}

\usepackage{url}

\usepackage{xcolor}

\usepackage{multirow}
\usepackage{graphicx}
\usepackage{subfig}
\usepackage{amsfonts}
\usepackage{bm}

\usepackage{amsmath,array}

\usepackage{booktabs}
\newcommand{\memo}[1]{\textcolor{blue}{[memo] #1}}
\usepackage{pifont}
\usepackage{colortbl}
\usepackage{xcolor}

\newcommand{\xmark}{\ding{55}}
\begin{document}
\maketitle

\begin{abstract}
Deep learning models have performed well on many NLP tasks. However, their internal mechanisms are typically difficult for humans to understand. 
The development of methods to explain models has become a key issue in the reliability of deep learning models in many important applications. 
Various saliency explanation methods, which give each feature of input a score proportional to the contribution of output, have been proposed to determine the part of the input which a model values most. 
Despite a considerable body of work on the evaluation of saliency methods, whether the results of various evaluation metrics agree with human cognition remains an open question.  
In this study, we propose a new human-based method to evaluate saliency methods in NLP by crowdsourcing. We recruited 800 crowd workers and empirically evaluated seven saliency methods on two datasets with the proposed method. We analyzed the performance of saliency methods, compared our results with existing automated evaluation methods, and identified notable differences between NLP and computer vision (CV) fields when using saliency methods. The instance-level data of our crowdsourced experiments and the code to reproduce the explanations are available at \url{https://github.com/xtlu/lreccoling_evaluation}.

\end{abstract}

\input{intro.tex}

\input{related.tex}

\input{proposed.tex}

\input{results.tex}

\input{conclusion.tex}


\section*{Acknowledgements and Ethics Statement}
This work was supported by JSPS KAKENHI Grant Number 23KJ1215, JSPS KAKENHI Grant Number 23H03402, JKA Foundation and JST CREST Grant Number JPMJCR21D1.

No personal information or data was used in this study. All crowd workers received appropriate wages in line with local standards.

\bibliography{ref}
\bibliographystyle{acl_natbib}

\newpage
\appendix
\onecolumn

\section{A demonstration of different saliency methods}

\begin{figure*}[t]
\centering
\centerline{\includegraphics[width=0.95\linewidth]{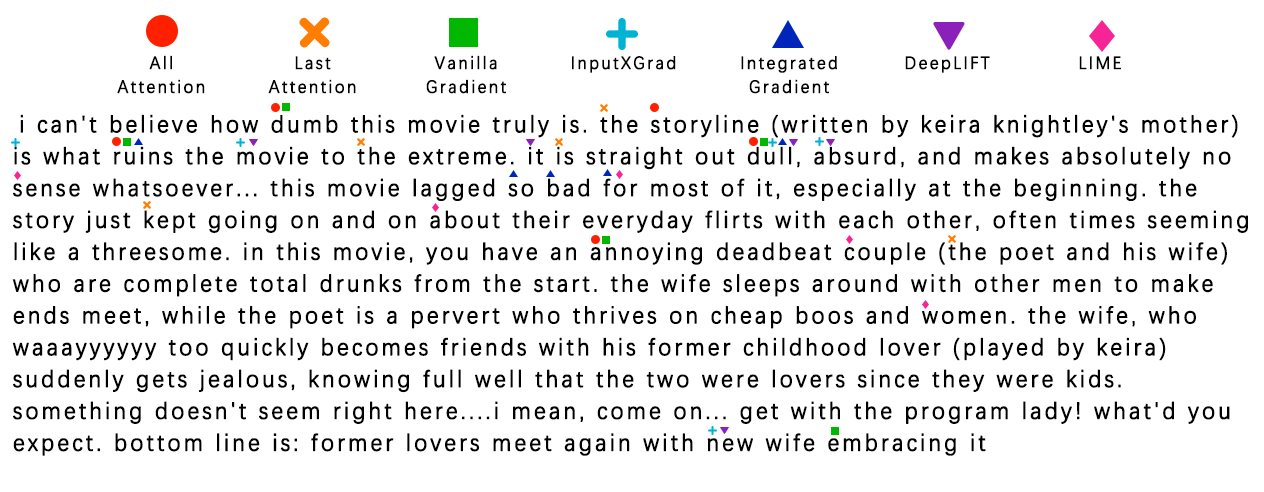}}
\caption{An example of the top five important words given by seven different saliency methods for a negative movie review in the IMDB dataset. A word with a mark (on the top of a word) represents as one of the top five important by a saliency method; marks with different colors and shapes represent different saliency methods. For example, the top five important words given by \textsc{LIME} consist of "sense", "woman", "for", "about", and "couple". The word "dull" is selected by five saliency methods. Even when some words are the same, different saliency methods are inconsistent with each other. }
\label{A1}
\end{figure*}

Figure~\ref{A1} shows the differing results (outputs) of different saliency methods for an example movie review. It could be found that LIME focused on some prepositions, such as "about" and "for" which are relatively meaningless for humans. 

\section{Task Instruction for Crowd Workers}
The instruction for the AGNEWS dataset is as follows, which is also similar to the instruction for the IMDB dataset. The original instruction was in Japanese. Even though Japanese native speakers have disadvantages in reading English text, we found that the quality of annotations collected in Lancers is higher than Amazon Mechanical Turk\footnote[5]{https://www.mturk.com/} empirically.

\begin{quotation}
[Read me first]

This task requires intermediate English reading ability.

We are collecting data on English news categories (“Worlds,” “Sports,” “Business,” and “Science” categories) for use in research on artificial intelligence and machine learning.

Information that can identify an individual or account, such as the user id of Lancers, will not be disclosed to third parties other than the task requester.

If you have any other questions, please contact us by message.

[Task instruction]

This task displays 100 different English news.
Classify each news item into one of the four categories ``Worlds'', ``Sports'', ``Business'' and ``Science''.

"Worlds" is news about world politics, crime, and conflict.

Example: "Guatemala to pay paramilitaries Guatemala's government agrees to pay millions of dollars to former paramilitaries, who some accuse of war crimes."

"Sports" is news about sports.

Example: "Iranian Will Not Meet Israeli in Olympics Due to Weight ATHENS (Reuters) - A diplomatic wrangle of Olympic proportions was avoided Sunday when Iranian world judo champion Arash Miresmaeili failed to make the weight for his clash with an Israeli opponent."

"Business" is news about finance, economy, and business.

Example: "SEC seeks to update stock sale rules BEIJING, Aug.18 (Xinhuanet) -- The US Securities and Exchange Commission (SEC) may revise its rules on initial stock offerings to modernize 1930s-era guidelines, agency commissioners and staff said. ”

"Science" is news about science, technology, and technology products.

Example: "India's leading IT company TCS to list stock next week after share offer (AFP) AFP - Top Indian software exporter Tata Consultancy Services will list its share on the Bombay Stock Exchange next week following its successful maiden share offer, an official said. ”

However, in the task, only a few selected words from the original news are presented. For example, if the original news is  "Israeli Attack in Gaza Kills Five Palestinians, Haaretz Says Aug. 18 (Bloomberg)".

What is presented is only a few discontinuous words, such as " . Attack .. Kills ... Says .. .. ".  ("." indicates a hidden word.)

Judge the class of news only from the words presented.

In this example, since the words "Attack" and "Kills" are included, it can be inferred that this news is probably related to world conflicts so that “Worlds” is its class.

If you cannot judge the class from the displayed words alone, select "I don't know". Also, if there are any English words you don't understand, you can translate them into Japanese using a translation website (for example, Google Translate).

This task can only be completed once per person. Even if you have multiple Lancers accounts, you can only do this once. That is all for the instruction. If there is no problem, please press the "Start" button.

\end{quotation}

\end{document}

%% file: intro.tex
\section{Introduction}

Recently, the development of deep neural network (DNN) models for natural language processing, such as BERT~\cite{devlin-etal-2019-bert}, and GPT-3~\cite{NEURIPS2020_1457c0d6}, has attracted considerable attention as a topic of active research. Given that DNNs can recognize important patterns in large and intricate datasets, DNN models exhibited significant achievements on various tasks for different applications. However, most deep models are black boxes, i.e., the relationship between their inputs and outputs is difficult to understand. 

In most low-risk tasks, errors made by DNNs do not have severe impacts; for example, in a recommendation system designed to suggest fiction novels, the impact of recommendation errors is relatively low. However, in some situations, the DNN model performing an NLP task could cause psychological harm to users. For example, an online chat service could output unethical or offensive sentences. To avoid this, one of the most fundamental and effective solution is to understand the decision-making process of the relevant models. Therefore, as the complexity of DNN models increases, the need for research on the interpretation of the output of such models becomes increasingly pressing.

Various interpretation methods have emerged to attempt to understand DNNs~\cite{liu-etal-2019-towards-explainable,moradi-etal-2021-measuring,carton-etal-2020-evaluating}, particularly for black-box predictions made by already-trained neural networks.
One of the major approaches to this problem involves the estimation of the influence of a subset of input features on the predictions of a model. 
By understanding the important features, the model can be improved, its predictions can be trusted more, and undesirable behaviors can be isolated~\cite{hooker2019benchmark}. Saliency explanations that output a contribution score for each feature in an input have been widely adopted in DNNs. Important features in input can be found by ranking contributions scores. Thus, saliency methods can be regarded as a general solution for the issue of black-box models, and various saliency methods have been proposed such as \textsc{Vanilla Gradient}s~\cite{vg1,vg2,vg3}, \textsc{Integrated Gradient}s~\cite{IG1,IG2}, and Local Interpretable Model-agnostic Explanations (\textsc{LIME})~\cite{ribeiro2016should}.


\begin{table*}[t]
    \centering
    \resizebox{1.0\linewidth}{!}{
    \begin{tabular}{lccccccc}
    \toprule

        &\textsc{All Attention} & \textsc{Last Attention} & \textsc{Vanilla Gradient} & \textsc{Input$\times$Grad}& \textsc{Integrated Gradient} & \textsc{DeepLIFT} & \textsc{LIME}\\
        \toprule
        \textsc{All Attention} & 100&	\cellcolor{cyan!10}50.4&\cellcolor{cyan!10}	57.5&	\cellcolor{cyan!30}31.2&	\cellcolor{cyan!30}34.8&	\cellcolor{cyan!30}31.8&\cellcolor{cyan!30}28.7\\

        \textsc{Last Attention} & \cellcolor{cyan!10}50.4&100&\cellcolor{cyan!10}48.5&\cellcolor{cyan!30}32.0&\cellcolor{cyan!30}33.9&\cellcolor{cyan!30}32.0&\cellcolor{cyan!30}29.0\\
        \textsc{Vanilla Gradient} &\cellcolor{cyan!10}57.5&	\cellcolor{cyan!10}48.5&	100&	\cellcolor{cyan!20}37.65&	\cellcolor{cyan!20}40.3&	\cellcolor{cyan!20}37.9&\cellcolor{cyan!30}28.8\\
        \textsc{Input$\times$Grad} & \cellcolor{cyan!30}31.2&	\cellcolor{cyan!30}32.0&	\cellcolor{cyan!20}37.65&	100&	\cellcolor{cyan!20}37.0&	\cellcolor{cyan!5}87.2&\cellcolor{cyan!30}28.3	\\
        \textsc{Integrated Gradient} &\cellcolor{cyan!30} 34.8&	\cellcolor{cyan!30}33.9&	\cellcolor{cyan!20}40.3&	\cellcolor{cyan!20}37.0&	100&	\cellcolor{cyan!20}37.5&\cellcolor{cyan!30}29.7 \\

        \textsc{DeepLIFT}& \cellcolor{cyan!30}31.8&\cellcolor{cyan!30}32.0&\cellcolor{cyan!20}37.9&\cellcolor{cyan!5}87.2&\cellcolor{cyan!20}37.5&100& \cellcolor{cyan!30}28.3 \\
        \textsc{LIME}& \cellcolor{cyan!30}28.7&\cellcolor{cyan!30}29.0&\cellcolor{cyan!30}28.8&\cellcolor{cyan!30}28.3&\cellcolor{cyan!30}29.7&\cellcolor{cyan!30}28.3&100  \\
        \bottomrule
    \end{tabular}
    }


       \caption{
       {\footnotesize
       \textbf{Overlap rate of seven different saliency methods based on top-40 words from 100 movie review samples.} For example, if we choose top-40 important words from \textsc{All Attention} and \textsc{LIME}, there are $28.7\% \times 40 = 11.48$ identical words on average. Value of 100 implies perfect overlap and \colorbox{cyan!40}{0.0} denotes no overlap.}}

\label{overlap}       
\end{table*}


The explanations rendered by various saliency methods often diverge; for example, we apply saliency methods on 100 movie review samples and found that the overlap rates, i.e. consistency of different saliency methods, were relatively low, as shown in Table~\ref{overlap}. Therefore, evaluating saliency methods becomes an important problem.

Along these lines, many automated evaluation metrics have been proposed to verify different interpretable models~\cite{madsen2021evaluating,hooker2019benchmark,pruthi2022evaluating,zhou2021evaluating}. However, such automated evaluation measures are not guaranteed to correctly evaluate the interpretability of saliency methods.
There are essential differences between machine cognition and human cognition. For example, \citet{su2019one} found that changing a single pixel in an image can change the output of models, whereas such changes are generally completely imperceptible to humans. Therefore, the interpretability of saliency methods should ultimately be decided by humans and a high rating by an automated evaluation metric does not necessarily mean a high interpretability for humans.
\citet{jacovi-etal-2023-neighboring} found that neighboring words could affect saliency explanations for human. \citet{de-langis-kang-2023-comparative} leveraged human eye-tracking to study saliency explanations. \citet{riordan2020analyzing} discussed human-machine agreements on different models for a gradient-based saliency method. However, evaluating saliency methods from the perspective of human intelligence has not been fully investigated.

Crowdsourcing provides an efficient and economical way to solve various tasks by utilizing human intelligence. 
For example, \citet{crowdsourcingtrust} proposed a mechanism of trust evaluation on edge computing using crowdsourcing. \citet{humanvision} exploited a crowdsourcing platform to evaluate human visual attention paths predicted by models.
Intuitively, we consider that a better saliency method should assign higher scores to more important words, enabling humans to correctly infer properties (e.g., categories) of the original text by the given several top important words. The key idea of this study is to apply crowdsourcing to evaluating saliency methods by whether crowd workers could distinguish labels of text based on the given top important words, as inspired by the measurements proposed by \citet{lu2021crowdsourcing} on image classification tasks. Based on the experimental results of the proposed evaluation method, we reach a list of findings on the saliency explanation methods in the NLP area. 

The contributions of this study are summarized as follows. 
  1) We propose a novel human-based method to evaluate saliency methods in NLP tasks by crowdsourcing. We recruited 800 workers from a crowdsourcing platform to evaluate seven saliency methods on two datasets and verified the reliability of the data of annotations provided by crowd workers. 2) We found that \textsc{Integrated Gradient}, a saliency method, performed best in our experiments and some saliency methods were not significantly better than random baseline. 3) We compared our results based on human evaluation with existing automated evaluation methods and found that there were differences between ours and existing methods. 4) We found a phenomenon called Flip, which makes saliency methods more unreliable in NLP than CV tasks. 5) We evaluated saliency methods on misclassified samples and found that saliency methods were also effective for misclassified samples.


%% file: related.tex
\begin{table*}[!t]
   \centering
    \resizebox{1.0\linewidth}{!}{
    \begin{tabular}{lccccccc}
    \toprule

        Method & \textsc{All Attention} & \textsc{Last Attention} & \textsc{Vanilla Gradient} & \textsc{Input$\times$Grad}& \textsc{Integrated Gradient} & \textsc{DeepLIFT} & \textsc{LIME}\\
        \midrule
        Type & Hidden States & Hidden States & Gradient-based & Gradient-based & Gradient-based & Gradient-based
        & Perturbation-based \\
       Class-specific &  \xmark  & \xmark    & \checkmark   & \checkmark   & \checkmark    & \checkmark    &  \checkmark \\
        \bottomrule
    \end{tabular}
    }

\caption{
{\footnotesize
\textbf{Saliency methods considered in this study.} "Class-specific" refers to whether a target class should be specified when using the method.}}

\label{methods}
\end{table*}

\section{Related Work}
\subsection{Saliency Methods}
We first briefly review the saliency methods used in this study. Table~\ref{methods} summarizes the saliency methods considered in this study. 

Saliency methods typically assign scores for each feature in an input, where important features with greater influence on the output are expected to receive a higher score. Most of the saliency methods can be divided into two types, including gradient-based and perturbation-based~\cite{slack2021reliable}.  Gradient-based saliency methods cannot be used for models that cannot be differentiable, such as random forest and perturbation-based saliency methods can be used in any model that maps an input to an output.

Gradient-based saliency methods assume that more important features have larger gradients. {\bf \textsc{Vanilla Gradient}}~\cite{vg1,vg2,vg3} is the most basic method, which measures the direct gradient of input features to quantify the influence of small changes in each input dimension on the output of the network. {\bf \textsc{Input$\times$Grad}}~\cite{gradxinput} multiplies the features in the input with the gradient with respect to the features. However, zero gradient does not necessarily mean zero contribution to the output. To solve this problem, {\bf \textsc{Integrated Gradient}}~\cite{IG1,IG2} is calculated by cumulating gradients of a straight-line path from given baselines to inputs. According to \citet{deeplift}, {\bf Deep Learning Important FeaTures (\textsc{DeepLIFT})} decomposes the contribution to the prediction output of a network into the contribution to each hidden neuron on a given input feature. 


Perturbation-based saliency methods assume that perturbing more important features exhibits larger influences on the output.
{\bf \textsc{LIME}}~\cite{ribeiro2016should} samples points around inputs and trains a simpler interpretable surrogate model, such as a linear model, based on the samples. The weights of the simple model are used as the contribution scores of inputs.

Except for common saliency methods, hidden states, i.e. variables or outputs of hidden layers, are sometimes explanations. In CV tasks, the output of convolutional layers may highlight important parts of the input image~\cite{zhou2016learning}. In this study, we define {\bf \textsc{All Attention}}~\cite{attentionmethod} as the mean of all layers and all heads of self-attention layers~\cite{vaswani2017attention} for a given word and {\bf \textsc{Last Attention}}~\cite{fernandes2022learning} as the mean of the last layer and all heads.

\subsection{Saliency Method Evaluation}
\label{secQ2}
We briefly introduce previous work of evaluating saliency methods. 

\textit{1) Relative Area Between-Curves (RACU): } 
\citet{madsen2021evaluating} utilized RemOve And Retrain (ROAR)~\cite{hooker2019benchmark} in NLP, which removes the top important words and retrains the model. The performance of saliency methods was measured in terms of the {\bf RACU}, which indicates changes in accuracy decline. 

\textit{2) Effectiveness (Eff.): }
\citet{pruthi2022evaluating} trained a student model with explanations of saliency methods of a teacher model and used {\bf Eff.} to measure how saliency methods helped student models train better. 

\textit{3) Simulability (Simu.): }
Similarly, \citet{fernandes2022learning} uses {\bf Simu.} to measure the simulation accuracy, i.e., what percentage of the student and teacher predictions matched over a dataset. 

\textit{4) Plausibility (Psi.) and Faithfulness (Fait.): }
\citet{ding-koehn-2021-evaluating} evaluated saliency methods from two perspectives, including {\bf Psi.} and {\bf Fait.} of model consistency on the PTB dataset~\cite{PTB}. {\bf Psi.} measures the similarity of explanations given by saliency methods and human annotations about the decision mechanism of a given model. {\bf Fait.} measures the consistency of the explanations given by a saliency method for the same input and two different models. 

\textit{5) Area Under the Precision-Recall curve (AUPRC): }
\citet{deyoung-etal-2020-eraser} collected annotations of words that humans think are important for decision-making and used {\bf AUPRC} to measure how well saliency methods agree with annotations marked by humans. 

\textit{6) Sufficiency (Suff.) and Comprehensiveness (Comp.)} were also proposed by \citet{deyoung-etal-2020-eraser} inspired by \citet{samek2016evaluating}.
{\bf Suff.} measures how model output confidence drops when the model only takes the important part given by a saliency method and assumes that confidence should drop slightly if the saliency method performs well. 
{\bf Comp.}, which is similar to ROAR but is not required to retrain the model, measures how model output confidence drops when the model takes the unimportant part, i.e., removing the important part from the original input, as the input and assumes that confidence should drop significantly if the saliency method performs well.

In this study, we evaluated saliency methods by humans directly. Crowd workers cannot see the full text while evaluating the important words given by saliency methods to make it easier for workers to make objective choices that are less dependent on subjective judgments of workers.

%% file: proposed.tex
\section{Crowd-based Evaluation Method}

\begin{figure}[!t]
\centering
\centerline{\includegraphics[width=1.0\linewidth]{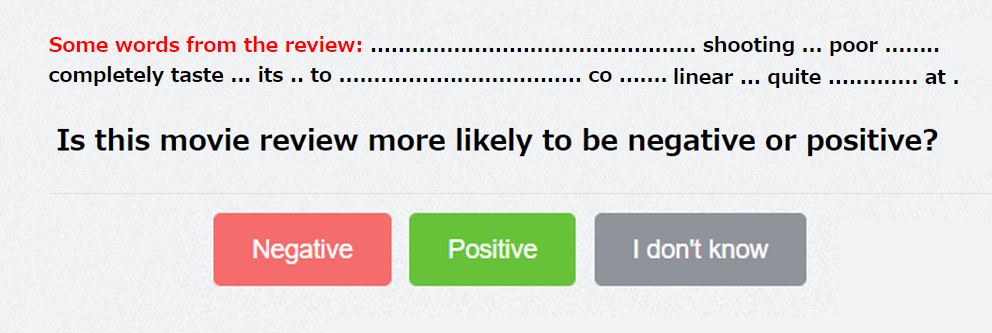}}
\caption{
{\footnotesize
\textbf{Example of a real task for crowd workers.} The top-$10$ words of a negative review given by a saliency method are shown. A "." corresponds to a hidden word. All punctuation and special tokens of the original text are ignored. Workers were asked to choose whether the review was likely to be negative, positive, or indeterminate based on the words shown. In this example, because the word "Poor" was included, this review may be inferred to be negative. Workers may obtain additional information or make inferences based on the position of words, which is consistent with the position encoding of the model. "."s prevent words that are far away from each other in the original text from being visually close, and will not mislead workers.
}
}
\label{interface}
\end{figure}

We evaluated saliency methods by showing the top important words given by the method to crowd workers.
In the web interface, crowd workers are asked to perform a text classification task, i.e., assigning a label to a text from a set of labels based on the words provided. Figure~\ref{interface} shows an example of a task.
We denote by $\mathcal{L}_{ijkm} \in \{-1,0,1,2,...,C\}$ the annotation given by crowd worker $m$ in the task $t_{ijk}$ of showing top-$k$ important words by the $j$-th saliency method for sample $i$, where $i \in \{1,2,...,N\}$, $N$ is the number of samples, $k \in \{\mathbb{Z^+}\}^K $, $K$ is the number of different top-$k$ words settings, and $C$ is the number of classes. $\mathcal{L}_{ijkm}=-1$ indicates that crowd worker $m$ did not evaluate task $t_{ijk}$, $\mathcal{L}_{ijkm}=0$ indicates that crowd worker $m$ selected "I don't know" in the task $t_{ijk}$ and $\mathcal{L}_{ijkm}=c \in \{1,2,...,C\}$ indicates that crowd worker $m$ selected label $c$ in the task $t_{ijk}$.

We denote by $\mathcal{A}_{ijk} \in \{0,1\}$ the results of whether task $t_{ijk}$ can be correctly recognized by crowd workers, where $\mathcal{A}_{ijk}=1$ means recognizable and vice versa.

We aggregate annotations from multiple workers by majority voting because of the large amount of tasks and the very sparse annotations with $\frac{3900}{4000}=97.5\%$ missing rate.
For a sample $i$ with ground truth label $c$, the aggregation can be represented by the following equation.
\begin{equation}
\mathcal{A}_{ijk} =\begin{cases}
  1&\text{if}\  \forall x,|\mathcal{L}_{ijkm}=c| > |\mathcal{L}_{ijkm}=x |,   \\
  0&\text{otherwise}. 
\end{cases}
\end{equation}
where $x\neq -1$ and $x\neq c$.

Next, we measure the performance of saliency methods in terms of the aggregation results $\mathcal{A}_{ijk}$. We consider that if a saliency explanation is sufficiently "interpretable", the accuracy, i.e., probability of labels can be correctly recognized more accurately by humans based on the given top-$k$ important words.

We denote by $p_{jk}=\frac{\sum_{i=1}^{N} \mathcal{A}_{ijk} }{N} $ the accuracy of the $j$-th saliency method for the top-$k$ words. The performance of the $j$-th saliency method $s_j$ can be considered as the weighted average of $p_{jk}$ of each $k$. When $k$ is relatively small, obtaining higher accuracy is more difficult than when $k$ is large, so we assign smaller $k$ larger weights. We use the mean accuracy of all saliency methods of the top-$k$ words as the difficulty of the top-$k$ words. The performance score can be represented as the following equation.
\begin{equation}
s_j=\sum_{k}^{} w_k p_{jk} \text{ , where } w_k = \frac{ \sum_{j}^{} \sum_{k}^{} p_{jk} }{ K^2 \sum_{j}^{}p_{jk}  }.
\end{equation}

The evaluation procedure consists of the following steps:

1) Prepare a pre-trained prediction model, a dataset, and several saliency methods to be evaluated. 

2) Apply all saliency methods and a random baseline to each text sample. 

3) Obtain saliency scores from each saliency method and the random baseline. 

4) Set the settings of top-$k$ important words. 

5) Prepare task $t_{ijk}$ of a given sample $i$, the $j$-th specified saliency method and show words $k$. 

6) Get annotations $\mathcal{L}_{ijkm}$ by crowdsourcing. 

7) Aggregate annotations $\mathcal{A}_{ijk}$ by majority voting.

8) Calculate performances $s_j$ of each saliency method.










%% file: results.tex
\section{Experiments}

We conducted experiments to answer the following four questions. 
1) Which saliency methods performed best in human evaluations? 
2) Do our results agree with the previous studies?
3) What is different between texts and images in the saliency methods?
4) What is the findings of evaluating misclassified cases?

\subsection{Experimental settings}
We used two datasets for the evaluation, including IMDB~\cite{IMDB} and AGNEWS, with a max length of $256$ tokens. IMDB is a binary text classification dataset that contains negative or positive movie reviews. AGNEWS is a subset of AG's corpus~\cite{AGNEWS}, which contains four classes of news items, including "World", "Sports", "Business" and "Science". 
We randomly selected 100 samples from both the IMDB and AGNEWS datasets of which 50 were negative movie reviews and 50 were positive in IMDB and 25 for each class in AGNEWS.
We used BERT-Base model~\cite{devlin-etal-2019-bert} as the backbone and trained classifiers by adding a linear layer to the BERT-Base model on the two datasets. The implementation of the BERT-base model was based on PyTorch and from HuggingFace~\footnote[1]{\url{https://huggingface.co/}}. 
All hyperparameters were set by default.
The top-$k$ important words were set as $5$, $10$, $20$, $30$, $40$ words, i.e., $k \in \{5,10,20,30,40\}$.
We evaluated the seven saliency methods in Table~\ref{methods} and a random baseline, i.e., assigning input features random contribution scores. The explanations of saliency methods were calculated based on Captum~\footnote[2]{\url{https://captum.ai/}} and all settings are default.

In total, 4000~\footnote[3]{100 samples $\times$ 8 (7 Saliency methods+Random baseline) $\times$ 5 different word settings = 4000.} tasks $t_{ijk}$ of (sample $i$, the $j$-th saliency method, top-$k$ words) were generated for each of two datasets. We used Lancers\footnote[4]{\url{https://www.lancers.jp/}} crowdsourcing platform for the evaluation. Each task was evaluated five times to collect reductant labels; the redundant labels are aggregated to ensure the reliability of the crowdsourcing results and eliminate the bias of individual workers.

Each crowd worker was required to evaluate 100 tasks for a reward of 330 JPY on Lancers. Each worker could only see each sample once to ensure that workers would not remember the content of the text. Therefore, cases that a worker completed a task with showing more words of a sample, then saw the same sample with fewer words and completed the task by the memory of the previous task would not happen.
400~\footnote[5]{4000 tasks * 2 datasets * 5 times / 100 tasks = 400. } crowd workers participated in the evaluations in total.

Workers completed the tasks in roughly 18 minutes on average. The hourly pay was about 1,100 JPY, which exceeds the minimum hourly wage, about 1,000 JPY, in Japan. Submissions of workers who completed the tasks within an extremely short period (less than 2 minutes) or showed obvious suspicions of inferior quality, such as selecting "I don't know" for all 100 tasks, were removed. No personal information or data was used in this study. All crowd workers received appropriate wages in line with local standards.

We randomly sampled 100 tasks out of 4,000 tasks in each dataset and let experts (members of our lab) evaluate these tasks to verify the reliability of the annotation data. 
The results show that the quality of the data collected was relatively reliable. The rate of agreement between experts and crowd workers, the rate of whether a task could be correctly recognized by both experts and crowd workers, was $86\%$ in IMDB dataset and $93\%$ in AGNEWS dataset. 

\begin{table*}[!t]

    \centering
    \resizebox{0.7\linewidth}{!}{
    \begin{tabular}{l|c|c|c|c|c|c}
    \toprule
        &5 Words & 10 Words & 20 Words & 30 Words& 40 Words & Score\\
        \midrule
        \textsc{Random} & 25 (8)&	42 (6)&41 (8)&	55 (6)&	54 (6)& 42.9 (7)\\
        \hline
        \textsc{All Attention} & \underline{66 (3)}&	\underline{74 (3)}&\underline{81 (2)}&	\textbf{\underline{87 (1)}}&	\underline{84 (3)}& \underline{78.5 (3)}\\

        \textsc{Last Attention} & 53 (4)&	60 (4)&\underline{73 (3)}&	72 (4)&	78 (4)& 66.9 (4)\\
        \textsc{Vanilla Gradient} & \textbf{\underline{75 (1)}}&	\textbf{\underline{81 (1)}}&\underline{73 (3)}&	\underline{80 (2)}&	\underline{85 (2)}& \underline{79.5 (2)} \\
        \textsc{Input$\times$Grad} & 46 (6)&	46 (5)&60 (5)&	56 (5)&	52 (7)& 52.1 (6)\\
        \textsc{Integrated Gradient} & \underline{74 (2)}&	\textbf{\underline{81 (1)}}&\textbf{\underline{87 (1)}}&	\underline{80 (2)}&	\textbf{\underline{87 (1)}}& \textbf{\underline{82.3 (1)}}\\

        \textsc{DeepLIFT} & 52 (5)&	37 (7)&60 (5)&	52 (7)&	67 (5)& 53.5 (5)\\
        \textsc{LIME} & 30 (7)&	28 (8)&47 (7)&	48 (8)&	44 (8)& 39.0 (8)\\
        \bottomrule
    \end{tabular}

    }

      \caption{
      {\footnotesize
      \textbf{Evaluation results of different saliency methods on the IMDB dataset.} The numbers in brackets show the ranks of the saliency method. The bold numbers show the best results and the underlined numbers show the top-$3$ methods. The results of top-$k$ words are the accuracy $p_{jk}$ (\%) and the last column "Score" are the overall performance $s_j$ (\%) of the $j$-th saliency method. Larger values indicate better performance.
      }}
    
\label{IMDBResults}
\end{table*}

\begin{table*}[!t]
   
    \centering
    \resizebox{0.7\linewidth}{!}{
    \begin{tabular}{l|c|c|c|c|c|c}
    \toprule
        &5 Words & 10 Words & 20 Words & 30 Words& 40 Words & Score\\
        \midrule
        \textsc{Random} & 46 (7)&	57 (8)&70 (5)&	77 (7)&	73 (8)&64.1 (7)\\
        \hline
        \textsc{All Attention} & \underline{73 (2)}&	\underline{73 (2)}&\underline{77 (2)}&	\underline{79 (3)}&	\underline{80 (2)}& \underline{76.9 (2)}\\

        \textsc{Last Attention} & 56 (6)&	\underline{72 (3)}&\underline{77 (2)}&	79 (3)&	\textbf{\underline{81 (1)}}& 72.8 (4)\\
        \textsc{Vanilla Gradient} & \underline{65 (3)}&	\underline{72 (3)}&69 (6)&	\underline{80 (2)}&	77 (4)& \underline{72.9 (3)} \\
        \textsc{Input$\times$Grad} & 61 (4)&	68 (5)&76 (4)&	78 (6)&	74 (7)& 71.5 (5)\\
        \textsc{Integrated Gradient} & \textbf{\underline{79 (1)}}&	\textbf{\underline{81 (1)}}&\textbf{\underline{87 (1)}}&	\underline{79 (3)}&	\underline{80 (2)}& \textbf{\underline{82.0 (1)}}\\

        \textsc{DeepLIFT} & 61 (4)&	61 (6)&69 (6)&	72 (8)&	77 (4)& 68.1 (6)\\
        \textsc{LIME} & 33 (8)&	61 (6)&69 (6)&	\textbf{\underline{82 (1)}}&	75 (6)& 62.9 (8)\\
        \bottomrule
    \end{tabular}
    }

      \caption{
       {\footnotesize
      \textbf{Evaluation results of different saliency methods on the AGNEWS dataset.} Same format with Table~\ref{IMDBResults}.
      }}

    \label{AGNEWSResults}
\end{table*}

\begin{table*}[!t]
\centering
\resizebox{1.0\linewidth}{!}{
\begin{tabular}{l|ccccccccc}
\toprule
Metrics & Ours$\uparrow$ &  RACU $\uparrow$   & Eff.$\uparrow$  & Simu.$\uparrow$  & Psi.$\uparrow$  & Fait.$\uparrow$ &  AUPRC $\uparrow$  &  Suff.$\downarrow$  & Comp.$\uparrow$   \\ \hline

 ~ & \multicolumn{9}{c}{\textit{Settings}}     \\ \hline

Model  & Bb & Rob   &  Bb  & Tf  & Tf  & Tf  & Bb+LSTM  & Bb  & Bb  \\ 

Dataset  & IMDB & IMDB   &  IMDB  & IMDB  & PTB  & PTB  & IMDB  & IMDB  & IMDB  \\ 

Human-included  & \checkmark & \xmark   &  \xmark  & \xmark   &\checkmark  & \xmark   & \checkmark & \xmark  &\xmark   \\ 
\midrule

 ~ & \multicolumn{9}{c}{\textit{Results}}                                                                                                                                                                                                    \\ \hline

Random               &  42.9(7)                & -                      & 91.5(7)                 & -                       & -                       & -                       & 0.259(4)                & 0.29(5)                & 0.04(7)                \\ \hline
\textsc{All Attention}        & \underline{ 78.5(3)}          & -                      & \textbf{\underline{94.55(1)}} & \textbf{\underline{86.27(1)}} & -                       & -                       & \textbf{\underline{0.417(1)}} & \underline{ 0.11(2)}          & \underline{ 0.28(2)}          \\
\textsc{Last Attention}       & 66.9(4)                & -                      & -                       & \underline{ 85.18(2)}          & -                       & -                       & -                       & -                      & -                      \\
\textsc{Vanilla Gradient}     & \underline{ 79.5(2)}          & \underline{ 25.4(2)}          & 92(4)                   & 83.14(4)                & \underline{ 0.551(2)}          & \underline{ 0.160(2)}          & \underline{ 0.385(2)}          & 0.25(4)                & 0.11(4)                \\
\textsc{Input$\times$Grad}          & 52.1(6)                & \underline{ 16.9(3)}          & 91.95(5)                & \underline{ 83.27(3)}          & -                       & -                       & -                       & 0.33(6)                & 0.06(5)                \\
\textsc{Integrated Gradient} & \textbf{\underline{82.3(1)}} & \textbf{\underline{35.1(1)}} & \underline{ 93.1(2)}           & 82.99(5)                & \textbf{\underline{0.734(1)}} & \textbf{\underline{0.239(1)}} & -                       & \underline{ 0.13(3)}          & \underline{ 0.17(3)}          \\
\textsc{DeepLIFT}             & 53.5(5)                & -                      & \underline{ 92.7(3)}           & -                       & -                       & -                       & -                       & 0.39(7)                & 0.06(5)                \\
\textsc{LIME}                 & 39.0(8)                & -                      & 91.95(5)                & -                       & -                       & -                       & \underline{ 0.280(3)}          & \textbf{\underline{0.06(1)}} & \textbf{\underline{0.32(1)}} \\
\bottomrule

\end{tabular}
}

\caption{
{\footnotesize
\textbf{Results of our method and previous studies on evaluating saliency methods.} $\uparrow$ indicates larger values are better and $\downarrow$ indicates smaller values are better. The bold numbers show the best results and the underlined numbers show top-$3$ methods. "Ours" refers to the overall performance scores of our proposed evaluation method. All abbreviations of metrics are defined in Section~\ref{secQ2}. All data are the original data reported in the previous studies. Especially, "Eff." and "Simu." are the average of different experiment settings. Data of "Comp." and "Suff."  were reported by \citet{pruthi2022evaluating} because of containing more saliency methods. "Bb", "Rob" and "Tf" are abbreviations of BERT-base, RoBERTa~\cite{liu2019roberta} and Transformer, respectively. }}

\label{tabel5}
\end{table*}

\subsection{Results and Discusisons}

\noindent
\textbf{Q1. Which saliency methods performed best in humans evaluations?  -- \textsc{Integrated Gradient}. }

Because different saliency methods provided different results of top-$k$ important words, we first investigated which saliency method yielded more reliable explanations for humans. It may be observed from Table~\ref{IMDBResults} and \ref{AGNEWSResults} that \textsc{Integrated Gradient} exhibited the best performance in the crowdsourced evaluation on both two datasets. This may be attributed to the fact that \textsc{Integrated Gradient} satisfies two axioms (desirable characteristics), including Sensitivity and Implementation Invariance~\cite{IG1}, whereas \textsc{Vanilla Gradient} as well as \textsc{Input$\times$Grad} violates the Sensitivity axiom and \textsc{DeepLIFT} violates the Implementation Invariance axiom.

\textsc{All Attention}, \textsc{Last Attention}, \textsc{Vanilla Gradient}, and \textsc{Integrated Gradient} performed significantly better than the random baseline on both two datasets; however, \textsc{Input$\times$Grad}, \textsc{DeepLIFT} and \textsc{LIME} did not. This most likely occurred because saliency methods cannot directly calculate the contribution scores for the words (tokens) in NLP tasks, but can calculate scores directly for pixels in CV tasks. Instead, the scores are calculated on token embeddings, and the calculation process may thus become more complicated and fail to achieve the desired performance.
\textsc{All Attention} performed better than \textsc{Last Attention} on both datasets, which indicates that each self-attention layer included valuable information.

Next, we discuss the impact of different datasets on the results. The rankings of saliency methods roughly remained the same on the two datasets with some differences. The results for IMDB are generally better than those for AGNEWS when few words were shown, although IMDB is a binary classification dataset and AGNEWS is a $4$-classification dataset. These results show that humans only need 5 words to achieve relatively high accuracy, but showing more words may not increase the accuracy for AGNEWS and \textsc{Integrated Gradient}. We discuss this further below.

\begin{table*}[!t]
\centering
\resizebox{0.9\linewidth}{!}{
\begin{tabular}{l|c|c}
\toprule

\multicolumn{1}{c|}{Text}                                                                                                                                                                                                          & $\#$ Words & Correct \\ \midrule
\begin{tabular}[c]{@{}l@{}}..........................................writing ............................................problem ................................\\ ............movie .........................insults ............................ruins .....................\end{tabular}                                                                                                                                                                 & 5                                                          & \checkmark     \\ \hline
\begin{tabular}[c]{@{}l@{}}...jason .....................................creative writing ............................................problem ...................\\ .........................movie .........................insults ............................ruins .life ..this .............movie ..\end{tabular}                                                                                                                                           & 10                                                        & \checkmark     \\ \hline
\begin{tabular}[c]{@{}l@{}}...jason .....................................creative writing .just ................crush .........................problem ..........\\ ..saying .........................aback .....movie.............course .........words .insults ...........tell ................ruins .\\ life.. this .the .......well ...movie .might\end{tabular}                                                                                     & 20                                                         & \xmark     \\ \hline
\begin{tabular}[c]{@{}l@{}}...jason .............built .....................finishes .creative writing .just .its .....fast ........crush .......................\\ ..problem .jason .......other ..saying .........................aback .....movie .............course ...believe .....words .insults .\\ jason .........tell ................ruins .life ..this .the short ...big ..well ...movie .might\end{tabular}                                         & 30                                                         & \xmark      \\ \hline
\begin{tabular}[c]{@{}l@{}}...jason .............built ..............cannot ......finishes .creative writing .just.its .....fast .hand ......crush ..............\\ ..........the problem .jason .......other way of saying .................movie .......aback .....movie .............course ...believe.\\ ....words .insults .jason ..go ......tell ................ruins .life ..this.the short ...big fat .well .a .movie it might\end{tabular}& 40                                                         &\xmark      \\ 

\bottomrule
\end{tabular}
}

\caption{
{\footnotesize
\textbf{An example of the top important words of a negative movie review provided by \textsc{Integrated Gradient}.} Humans correctly recognized that it is negative when showing $5$ and $10$ words, but failed when showing $20$, $30$ and $40$ words. It is probably because the words "well" and "movie" affect human judgment.}}

\label{textcase}
\end{table*}

\noindent
\textbf{Q2. Do our results agree with the previous studies? -- Some agree and some disagree. }

We compare the results of previous evaluation methods mentioned in Section \ref{secQ2} with ours in Table~\ref{tabel5}. Although models, datasets, and experiment settings in these works differ from each other somewhat, findings can be summarized as follows.

1) These results show that RACU, Psi. and Fait. completely agree with our results, but the results of other automated metrics were not all completely the same as our results. Psi. and AUPRC include human participation but are not directly human evaluations. The differences between our results and AUPRC may be attributed to the fact that the spans or sentences marked by human annotations were not very accurate or rankings of words were not provided.
It is also possible that the reason that these three metrics agree with ours completely was that relatively few saliency methods were included in the experiments.

2) \textsc{Integrated Gradient} performed best among gradient-based saliency methods, and \textsc{Vanilla Gradient} performed better than \textsc{Input$\times$Grad}, except for Simu. metric. 

3) Comp. and Suff. show that \textsc{LIME} performed best, which was the opposite of our results. These findings show that saliency methods that perform well in terms of automated evaluation metrics do not necessarily perform well in terms of human evaluation.

\noindent
\textbf{Q3. What is the difference between the results of the saliency methods for texts and images? -- The Flip phenomenon.}

\begin{figure}[!t]
\centering
\centerline{\includegraphics[width=1.8in]{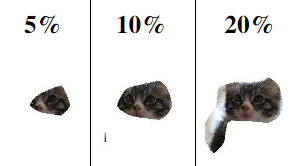}}
\caption{
{\footnotesize
\textbf{An example~\cite{lu2021crowdsourcing} of top-$5\%$ / $10\%$ / $20\%$ important pixels given by vision saliency method GradCAM~\cite{selvaraju2017grad}.} Regardless of the saliency method, once humans can recognize that the images with fewer pixels show a cat, it is almost impossible for us to fail to recognize the same cat image with more pixels. However, there are two exceptions. The first is that the image is ambiguous. For example, in the cat/dog classification problem, an image may include both a cat and a dog. Another issue is rarer, that is, the position of important pixels can express the outline of an object's shape, such as by showing an outline of a cat's head against a gray sky~\cite{gupta2022new}.}}
\label{gradcam}
\end{figure}

Most studies on saliency methods were first introduced in CV. Finally, we discuss a phenomenon called {\bf Flip} which occurs frequently in NLP but is rarely considered in CV. A sample $i$ is Flipped with the $j$-th saliency method when humans can correctly distinguish the label of a sample $i$ when showing fewer words/pixels, but can not distinguish with showing more words/pixels, i.e. $\forall k_1 \exists k_2>k_1 ,\mathcal{A}_{ijk_1}=1  \text{ and } \mathcal{A}_{ijk_2}=0$. We show a text example in Table~\ref{textcase} and an image example in Figure~\ref{gradcam}, respectively.

\begin{figure}[!t]
\centering

\subfloat[IMDB]{
            \includegraphics[width=0.48\linewidth]{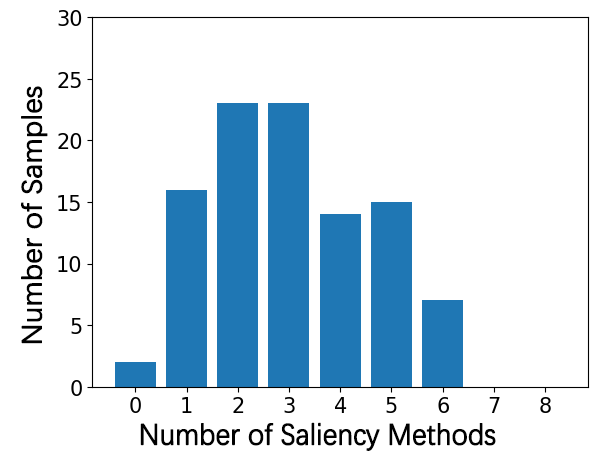} } 
\subfloat[AGNEWS]{\includegraphics[width=0.48\linewidth]{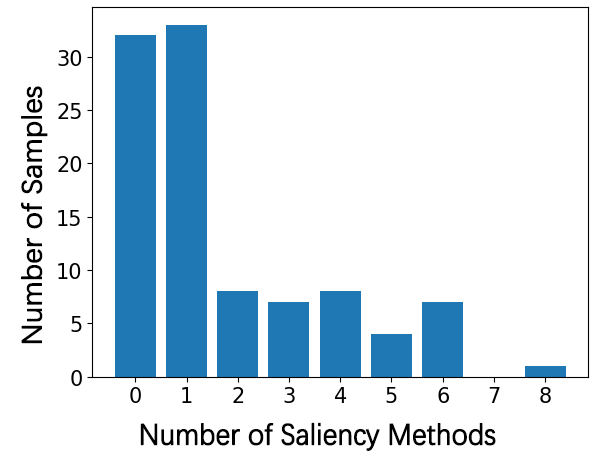} }

\caption{
{\footnotesize
\textbf{Histograms of Flips in the two datasets.} The horizontal and vertical axes indicate the number of saliency methods where a sample Fliped with and the number of samples, respectively.  For example, over 30 samples among 100 samples in the AGNEWS dataset did not Flip with all the $8$ saliency methods (including the random baseline). Regardless of saliency methods, some samples are easier to Flip, i.e., contain more misleading words. 
}}
\label{hist}
\end{figure}

We consider that the Flip phenomenon is saliency method-agnostic but attribute this to the properties of words themselves. Unlike pixels, each word has the potential to have strong expressive power to affect the meaning of the entire sample. Figure~\ref{hist} indicates that some samples were inclined to be Flipped and Table~\ref{flip} indicates that human recognition is not necessarily aided by showing more words. We noticed that Flip appeared more in the IMDB dataset, possibly because the sentiment of movie reviews was more easily influenced by individual words. Based on these findings, we suggest that future research on the application of saliency methods in NLP may consider the Flip phenomenon. 

\begin{table}[t]

    \centering
    \resizebox{1.0\linewidth}{!}{
    \begin{tabular}{l|cc|cc}
    \toprule
        \multirow{2}{*}{~} & \multicolumn{2}{c|}{IMBD} & \multicolumn{2}{c}{AGNEWS}  \\
        &Flips&Aids&Flips&Aids \\
        \midrule
        \textsc{Random} & 42&	46&19&	41\\
        \hline
        \textsc{All Attention} & 18&	27&20&	19\\

        \textsc{Last Attention} & 27&36&10&	39\\
        \textsc{Vanilla Gradient} & 31&	18&29&	26\\
        \textsc{Input$\times$Grad} & 54&	33&28&	26\\
        \textsc{Integrated Gradient} & 28&	20&15&	15\\

        \textsc{DeepLIFT} & 50&	34&23&	33\\
        \textsc{LIME} & 54&	42&23&	54\\
        \bottomrule
    \end{tabular}
    }

     \caption{
     {\footnotesize
     \textbf{Numbers of Flipped samples among 100 samples in two datasets.} "Aids" means that showing more words aids humans in correctly recognizing the label. That is $100-$$\#$ Flips$-\#$ Samples that can be correctly recognized in all the top-words settings.}}
   
    \label{flip}
\end{table}

\begin{table*}[!t]
    \centering
    \resizebox{0.64\linewidth}{!}{
    \begin{tabular}{l|c|c|c|c|c}
    \toprule
        &5 Words & 10 Words & 20 Words & 30 Words& 40 Words \\
        \midrule
        \textsc{Random} & 19 &	45  & 43 &	49  &  38  \\
        \hline
        \textsc{All Attention}   & 15 &	25  & 41 &	35  &  38  \\
        \textsc{Last Attention}  & 12 &	32  & 35  &	40  &  37  \\
        \textsc{Vanilla Gradient}  & 33 &	39  & 37 &	31  &  50  \\  
        \textsc{Input$\times$Grad}   & 28 &	20  & 45 &	35  &  43  \\ 
        \textsc{Integrated Gradient} & 21 &	44  & 31 &	38  &  30  \\  
        \textsc{DeepLIFT}  & 29 &	25  & 33 &	44  &  34  \\ 
        \textsc{LIME}  & 14 &	33  & 36 &	35  &  34  \\
        \bottomrule
    \end{tabular}
    }

       \caption{
       {\footnotesize
       \textbf{Evaluation results of misclassified cases in the IMDB dataset.} The number is the count of samples which could be recognized by humans correctly from 100 samples which were all misclassified by the model.}}

 \label{IMDBwrong}       
\end{table*}

\begin{table*}[!t]

    \centering
    \resizebox{0.64\linewidth}{!}{
    \begin{tabular}{l|c|c|c|c|c}
    \toprule
        &5 Words & 10 Words & 20 Words & 30 Words& 40 Words \\
        \midrule
        \textsc{Random} & 15 &	20  & 26 &	23  &  24  \\
        \hline
        \textsc{All Attention}   & 24 &	22  & 16 &	30  &  26  \\
        \textsc{Last Attention}  & 22 &	25  & 21  &	25  &  23  \\
        \textsc{Vanilla Gradient}  & 29 &	21  & 23 &	25  &  23  \\  
        \textsc{Input$\times$Grad}   & 26 &	30  & 27 &	20  &  22  \\ 
        \textsc{Integrated Gradient} & 50 &	40  & 37 &	34  &  26  \\  
        \textsc{DeepLIFT}  & 28 &	31  & 21 &	22  &  21  \\ 
        \textsc{LIME}  & 18 &	22  &  23 &	25  &  24  \\
        \bottomrule
    \end{tabular}
    }

    \caption{
    {\footnotesize
    \textbf{Evaluation results of misclassified cases in the AGNEWS dataset.} The format is the same as Table~\ref{IMDBwrong}. }}

    \label{AGNEWSwrong}
\end{table*}

\noindent
\textbf{Q4. What is the findings of evaluating misclassified cases? -- Effective for explaining models. }

In the previous experiments, we followed the settings that are usually used in the existing works~\cite{deyoung-etal-2020-eraser,samek2016evaluating}, in which all chosen data samples were predicted by the model correctly, because saliency methods usually could not perform well on misclassified samples.
However, misclassified cases are also important because analyzing misclassified cases is beneficial for understanding the model better. 
Therefore, we also conducted the experiments with misclassified cases by randomly choosing $100$ misclassified samples from each dataset, and recruited $400$ crowd workers to conduct the same experiments, as shown in Table~\ref{IMDBwrong} and \ref{AGNEWSwrong}. The findings of the experimental results are list as follows.

1) Comparing with Table~\ref{IMDBResults} and \ref{AGNEWSResults}, the performance of the random baseline has dropped significantly, which indicates the inherent difficulty of misclassified samples are relatively high and more ambiguous regardless of the model.

2) The performance of all methods also drops significantly, indicating that the correct top important words for correct prediction are not highlighted by the model. Therefore, saliency methods properly present the issue of misclassifications.

3) Showing more words does not necessarily improve the performance. This phenomenon is also present in the correct classified samples shown in Q3, but there are more ambiguous words in misclassified samples, which may even reduce the overall accuracy.

4) Especially, \textsc{Integrated Gradient} obtained $50\%$ accuracy and outperforms other saliency methods significantly when showing top $5$ words on the AGNEWS dataset; however, the low performance of \textsc{All Attention} indicates that the self-attention layers do not give high weights to the important words. Misclassification may tend not to occur when \textsc{All Attention} performs well because attentions are based on the self-attention layers of the model. The low performance of \textsc{Vanilla Gradient} indicates that the model is not currently highlighting top important words, i.e. the model is not currently learning effectively on misclassified samples. Compared with \textsc{Vanilla Gradient}, the reason why \textsc{Integrated Gradient} performs well when showing $5$ words may be attributed to that \textsc{Integrated Gradient} has the property of completeness, which may highlight potential top important words that can be learned by the model.

%% file: conclusion.tex
\section{Conclusion}
In this study, we proposed an approach to evaluate saliency explanations in natural language processing by crowdsourcing.
The proposed approach assesses the interpretability of saliency methods in terms of human evaluations. Based on this approach, We recruited a total of 800 crowd workers to conduct our experiments for evaluating seven different saliency methods on two datasets.

The scale of our experiments was limited by the cost of crowdsourcing because real crowdsourcing experiments are costly and adding more datasets or models will dramatically increase the cost. Although the number of datasets is relatively small and the results were only evaluated on the BERT-base model, the main purpose of this work is to claim that there are differences between human evaluation and automated evaluations and there are indeed differences in our experiments of one model and two datasets. As long as there are differences in one model, there are differences. Although more experiments will support our claim, we would like to caution further studies when using automated evaluation methods. Four takeaways can be summarized as follows based on the results. 

1) \textsc{Integrated Gradient} performed best in our experiments. \textsc{All Attention}, \textsc{Last Attention} and \textsc{Vanilla Gradient} performed significantly better than the random baseline on both two datasets. \textsc{DeepLIFT} and \textsc{LIME} performed close to or worse than the baseline, probably because the contribution scores were calculated on token embeddings but not directly on tokens (words), which leads to additional complexity in the calculation.

2) There were differences between our results based on human evaluation and the results based on some of the existing automated evaluation metrics. Except for our results, \textsc{Integrated Gradient} also performed best among Gradient-based saliency methods in most evaluation metrics. 

3) A phenomenon called Flip occurred in NLP which is saliency method-agnostic and rarely applied in CV. We also found that showing more words to humans may not help them correctly recognize the labels of a given text. In future work, we will research how to avoid the Flip phenomenon of saliency methods in NLP.

4) The reasons for misclassification can be attributed to two perspectives. The first is that the inherent difficulty of misclassified samples is relatively high according to the performance of the random baseline. The other is that the model failed to highlight the correct top important words according to the poor performances of saliency methods. 
